\documentclass[aps,prd,twocolumn,nofootinbib,preprintnumbers,amssymb,amsfonts,amsmath,superscriptaddress,showpacs]{revtex4-1}

\usepackage{graphicx}
\usepackage{bm}
\usepackage{hyperref}
\usepackage{color}
\usepackage{subfigure}
\usepackage{url}
\usepackage{epstopdf}
\usepackage{multirow}
\usepackage{enumerate}

\def\simleq{\; \raise0.3ex\hbox{$<$\kern-0.75em \raise-1.1ex\hbox{$\sim$}}\; }
\def\simgeq{\; \raise0.3ex\hbox{$>$\kern-0.75em \raise-1.1ex\hbox{$\sim$}}\; }

\begin{document}

\title{
Gamma-ray sky points to radial gradients in cosmic-ray transport
}

\author{Daniele Gaggero}
\email{daniele.gaggero@sissa.it}
\affiliation{SISSA, via Bonomea 265, I-34136, Trieste, Italy}
\affiliation{INFN, sezione di Trieste, via Valerio 2, I-34127, Trieste, Italy}

\author{Alfredo Urbano}
\email{alfredo.urbano@sissa.it}
\affiliation{SISSA, via Bonomea 265, I-34136, Trieste, Italy}

\author{Mauro Valli}
\email{mauro.valli@sissa.it}
\affiliation{SISSA, via Bonomea 265, I-34136, Trieste, Italy}
\affiliation{INFN, sezione di Trieste, via Valerio 2, I-34127, Trieste, Italy}

\author{Piero Ullio}
\email{piero.ullio@sissa.it}
\affiliation{SISSA, via Bonomea 265, I-34136, Trieste, Italy}
\affiliation{INFN, sezione di Trieste, via Valerio 2, I-34127, Trieste, Italy}



\begin{abstract}
The standard approach to cosmic-ray (CR) propagation in the Galaxy is based on the assumption that local transport properties can be extrapolated to the whole CR confining volume. Such models tend to underestimate the $\gamma$-ray flux above few GeV measured by the Fermi Large Area Telescope towards the inner Galactic plane. We consider here for the first time a phenomenological scenario allowing for both the rigidity scaling of the diffusion coefficient and convective effects to be position-dependent. We show that within this approach we can reproduce the observed $\gamma$-ray spectra at both low and mid Galactic latitudes -- including the Galactic center -- without spoiling any local CR observable.
\end{abstract}

\maketitle

\section{Introduction}

Since 2008 the Fermi Large Area Telescope ({\tt Fermi-LAT}) has been surveying the $\gamma$-ray sky between about few hundred MeV and few hundred GeV with unprecedented sensitivity and resolution. The bulk of the photons detected by the {\tt Fermi-LAT} is believed to be associated with diffuse emission from the Milky Way, originated by Galactic cosmic rays (CRs) interacting with the gas and the interstellar radiation field (ISRF) via production and decay of $\pi^0$s,  inverse Compton (IC), and bremsstrahlung. 

There is a striking consistency between general features in the diffuse $\gamma$-ray maps and the diffuse $\gamma$-ray flux models: the predictions mainly rely, on the side concerning emitting targets, on (indirectly) measured gas column densities and ISRF models, while, on the side of incident particles, on propagation models tuned to reproduce locally measured fluxes. When addressing at a quantitive level the quality of such match between predictions and data, most analyses have mainly developed  optimized models looping over uncertainties on the emitting targets. In particular, in ref.~\cite{FermiLAT:2012a} the authors -- besides allowing for a radially-dependent rescaling of the ISRF and different values of the spin temperature of the 21 cm transition -- adopt a tuning of the poorly known conversion factor between the observed CO emissivities and the molecular hydrogen column densities, usually dubbed ${\rm X_{CO}}$. In ref.~\cite{FermiLAT:2012a} it is shown that such approach is sufficient to generate models in agreement with the data within about $15$\% in most regions of the sky; a remarkable exception is the fact that this procedure tends to systematically underestimate the measured flux above few GeV in the Galactic plane region, most notably towards the inner Galaxy. 

\begin{figure}[!h!]
\vspace{-0.2cm}
\centering
   \includegraphics[scale=0.45]{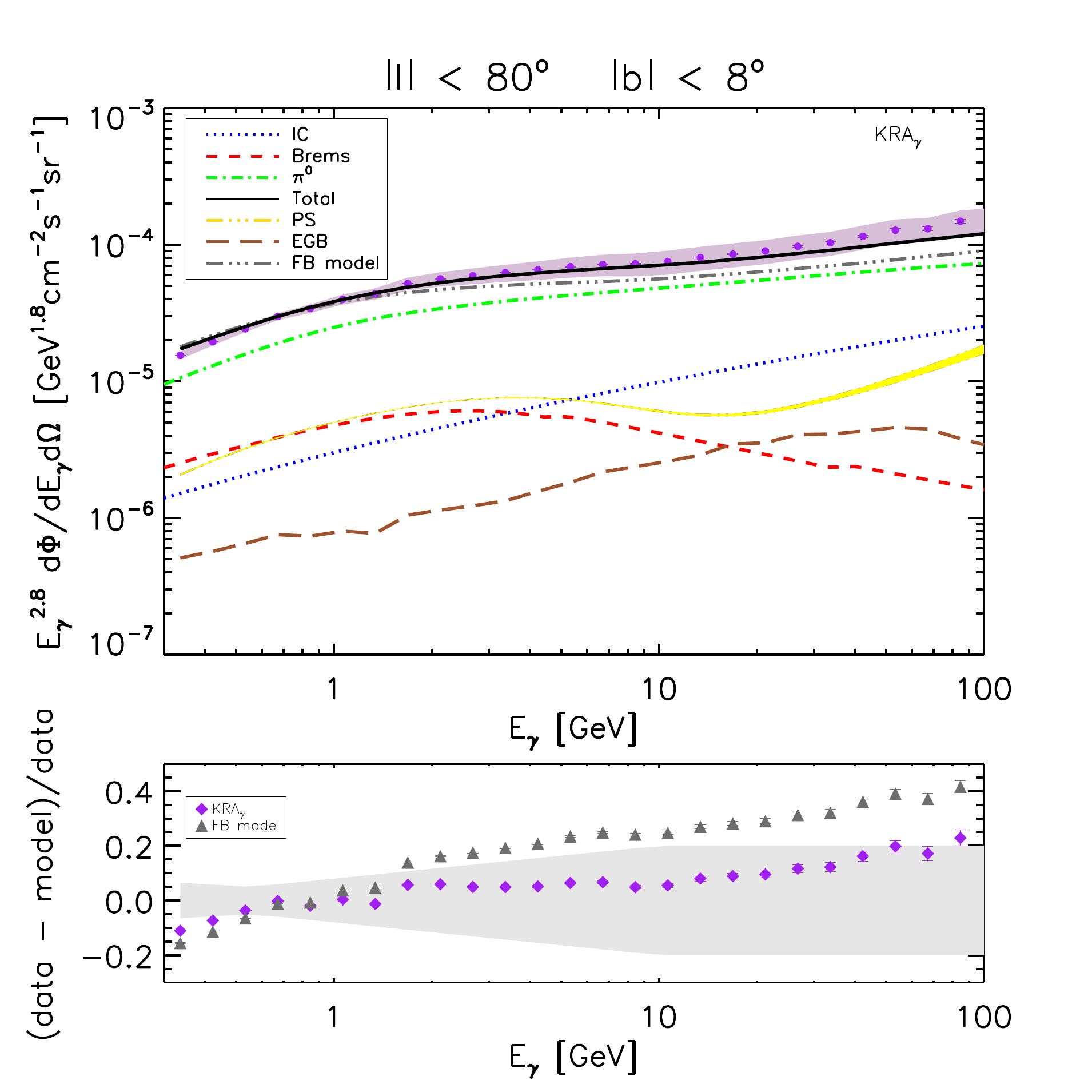}
\caption{\textit{
{Upper panel. Comparison between the $\gamma$-ray flux computed with the CR propagation model proposed in this Letter (KRA$_{\gamma}$ total flux: solid black line; individual components shown) and the {\tt Fermi-LAT} data (purple dots, 
including both statistic and systematic errors)
in the Galactic disk. For comparison, we also show the total flux for the FB model defined in ref.~\cite{FermiLAT:2012a} (double dot-dashed gray line). Lower panel. Residuals computed for the KRA$_{\gamma}$ and FB models.} 
}}
\label{fig:Spectra_KRA_tuned_plane}
\end{figure}

Fig.~\ref{fig:Spectra_KRA_tuned_plane} shows the spectrum for the $\gamma$-ray flux measured by the {\tt Fermi-LAT} in the energy range between 300~MeV and 100~GeV and a large angular window encompassing the inner Galactic plane ($5$ years of data, within the event class {\tt ULTRACLEAN} according to Fermi tools {\tt v9r32p5}, as described in \cite{WeiEAlfredo}). The yellow band corresponds to the point sources (PS) modelled using the 2-years {\tt Fermi-LAT} Point Source Catalogue via a dedicated Monte Carlo (MC) code. The brown line is the contribution of the  extragalactic background (EGB) obtained by a full-sky fit of the data for $|b| > 20^{\circ}$. The  double dot-dashed line and gray triangles are, respectively, the prediction and residuals for the Fermi benchmark model, labelled $^{\rm S}$S$^{\rm Z}$4$^{\rm R}$20$^{\rm T}$150$^{\rm C}$5 (FB hereafter), selected for fig.~17 in ref.~\cite{FermiLAT:2012a}, and reproduced here using the 
{\tt GALPROP} WebRun \citep{Galpropwebrun,Vladimirov2011}: while the model is optimized at low energy, it gives a poorer description of the data at high energy, a feature that is generic for all models proposed in that analysis.    

The selected angular window is interesting because the diffuse emission from the inner Galactic plane is potentially a precious source of information for CR transport modelling. Being the region with largest gas column densities, it is the brightest zone of the sky and, unlike other regions where the interplay among components allows more modelling freedom, its flux is predominantly shaped by only one contribution, namely the $\pi^0$ decays, especially when looking at intermediate energies. The $\pi^0$ emissivity spectral index is roughly equal to the incident proton one, hence the inner Galactic plane allows an indirect measurement of the CR proton slope towards the center of the Galaxy, far away from the region where direct measurements are available. This aspect is seldom emphasized, since the standard approach consists in solving the propagation equation for CR species \citep{Berezinsky_book} under the assumption that diffusive properties of CRs are the same in the whole propagation volume. This implies reducing the spatial diffusion tensor to a single constant diffusion coefficient $D(\rho) = D_0 (\rho/\rho_0)^{\delta}$, whose scaling $\delta$ on rigidity $\rho$ and normalization $D_0$ are constrained by local CR data (a range between about $\delta = 0.3$ and about $\delta = 0.85$ is allowed \citep{2001ApJ...555..585M,DiBernardo:2009ku,Trotta:2010mx}). Such hypothesis freezes the proton spectral index -- and therefore the $\pi^0$ spectral index -- to be very close to the local one everywhere in the CR propagation region. For this reason, in fig.~\ref{fig:Spectra_KRA_tuned_plane} and in the following, the $\gamma$-ray flux is multiplied by $E_{\gamma}^{2.8}$, since $\gamma_p=2.820\pm0.003 \,{\rm(stat)}\,\pm0.005  \,{\rm(sys)}$ is the proton index measured by the PAMELA experiment in the range 30~GV--1.2~TV~\cite{Adriani:2011cu}. The FB model gives a slightly rising curve since it assumes $\gamma_p = 2.72$. 

%
%

The present analysis goes beyond standard approaches by allowing for spatial gradients in diffusion,
using as a guideline the {\tt Fermi-LAT} $\gamma$-ray data. 

In the CR transport equation, the diffusion term describes at macroscopic level the effective interplay between CRs and the magnetohydrodynamics turbulence, see, e.g., ref.~\cite{Schlickeiser2002}. In the framework of quasi-linear theory (QLT), $\delta$ is related to the turbulence spectrum (e.g. $\delta=1/3$ for Kolmogorov-like turbulence and $\delta=1/2$ for Kraichnan-like one); QLT however assumes that the turbulent component of the magnetic field is subdominant compared to the regular one, an hypothesis that does not seem to be supported by recent models~\cite{Jaffe2011,Jansson:2012pc}. Studies based on non-linear theory approaches, on the other hand, find more involved environmental dependencies, resulting in different scalings in different regions of the Galaxy, and deviations from a single power law in rigidity~\cite{Yan2008,Evoli2014}. 
An additional element to take into account is the possibility that CRs themselves generate the turbulent spectrum responsible for their propagation~\cite{BlasiSerpico:2012}, introducing local self-adjustments in propagation.  

Given these arguments, in the following we will consider models with variable $\delta$ and show how they naturally improve the description of $\gamma$-ray data.

\begin{table}[!htb!]
\centering
\begin{tabular}{|c|c||c|c|}
\hline
sky window  &   $\alpha$ & sky window  &   $\alpha$  \\
   ($|b| < 5^{\circ}$)   &  $(\Phi \sim E_{\gamma}^{-\alpha})$  &   ($|b| < 5^{\circ}$)  & $(\Phi \sim E_{\gamma}^{-\alpha})$   \\
\hline
{\,\,\,$0^{\circ}<|l|<10^{\circ}$}  & {$2.55\pm0.09$}  & {$40^{\circ}<|l|<50^{\circ}$}  &  {$2.57\pm0.09$}\\   
{$10^{\circ}<|l|<20^{\circ}$}  &  {$2.49\pm0.09$} & {$50^{\circ}<|l|<60^{\circ}$}  &  {$2.56\pm0.09$}\\
{$20^{\circ}<|l|<30^{\circ}$}  & {$2.47\pm0.08$} & {$60^{\circ}<|l|<70^{\circ}$}  &  {$2.60\pm0.09$}\\
{$30^{\circ}<|l|<40^{\circ}$}  &  {$2.57\pm0.08$} & {$70^{\circ}<|l|<80^{\circ}$}  &  {$2.52\pm0.09$}\\
\hline
\end{tabular}\\[0.1cm]
\caption{\label{tab:slopes} \textit{
Energy slope of {\tt Fermi-LAT} $\gamma$-ray data on the Galactic disk. The 
power-law index has been obtained by fitting the data in the energy window $E_{\gamma} = [5-50]$ GeV.
We average in latitude over the interval $|b| < 5^{\circ}$.
}}
\end{table} 

\begin{table*}[!htb!]
\centering
\begin{tabular}{|c|c|c|c|c|c|}
\hline
$\chi^2$ values &  $0^{\circ} < |l| < 80^{\circ}$ & $0^{\circ} <  |l| < 10^{\circ}$ & $20^{\circ} < |l| < 30^{\circ}$ & $50^{\circ} < |l| < 60^{\circ}$ & ~~\,$0^{\circ} < |l| < 180^{\circ}$ \\ 
($25$ data points)  & $0^{\circ} < |b| < 8^{\circ}$~\, & $0^{\circ} < |b| < 5^{\circ}$~\, & $0^{\circ} < |b| < 5^{\circ}$~\, &  $0^{\circ} < |b| < 5^{\circ}$  & $10^{\circ} < |b| < 20^{\circ}$ \\
\hline
{$\chi^2$  KRA$_{\gamma}$}  & {11.30} & {3.79}  & {12.27} & {11.50} & {6.94} \\
{$\chi^2$ FB model}  & {53.00} & {74.83} & {70.04}  & {24.85} & {17.60} \\
\hline
\end{tabular}\\[0.1cm]
\caption{\label{tab:chi2} \textit{Results of the $\chi^2$ analysis for the fit of the {\tt Fermi-LAT} $\gamma$-ray data. }}
\end{table*}   

\section{Analysis.} 

{We decide to follow a data-driven approach.
In order to quantify the change of the $\gamma$-ray slope along the Galactic disk and the resulting discrepancy between the FB model and the actual data, we show in 
table~\ref{tab:slopes} the power-law index obtained by fitting the {\tt Fermi-LAT} $\gamma$-ray data in the energy window $E_{\gamma} = [5-50]$ GeV, and in the second row of table~\ref{tab:chi2} the $\chi^2$ of the FB model. 
 
{ The observed power-law
index ranges from $E_{\gamma}^{-2.47}$ to $E_{\gamma}^{-2.60}$, thus resulting in a $\gamma$-ray flux much harder than the prediction of the FB model, especially in the central windows.
These data should be taken as a guideline, and only show a hint of a slope change with $l$, instead of a statistically robust evidence. We remark that, in the outermost windows we considered, the gamma-ray emission is not dominated by $\pi^0$ emission only, since the relative contributions of point sources and Inverse Compton are far from being negligible. 
}

\begin{figure}[!htb!]
\centering
   \includegraphics[scale=0.4]{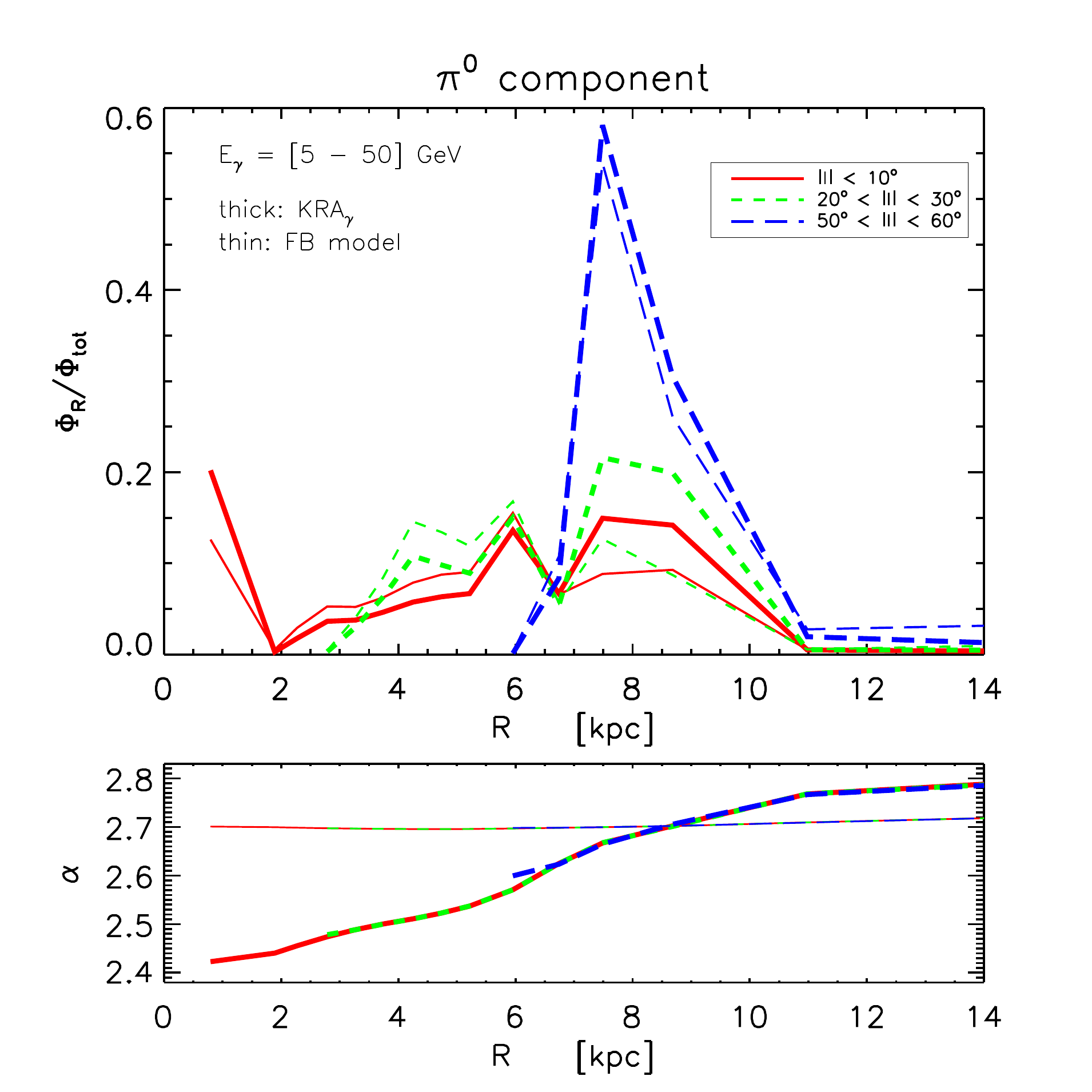}
\caption{\textit{
{ 
Relative contribution (upper panel),
and power-law spectral index of the $\pi^0$ emission (lower panel, with scaling $\sim E_{\gamma}^{-\alpha}$)
for three reference l.o.s.
as a function of the radial distance from the Galactic center. The FB (KRA$_{\gamma}$) model corresponds to thinner (thicker) lines. 
We average in latitude over the interval $|b| < 5^{\circ}$.
}
}}
\label{fig:Ring_Spectrum}
\end{figure}

Turning our attention to the quality of the fit for the FB model is worse in the innermost windows (e.g. $|l| < 10^{\circ}$ and $20^{\circ} <  |l| < 30^{\circ}$, 
with $|b| < 5^{\circ}$),
it slightly ameliorates going towards outer longitudinal values ($50^{\circ} <  |l| < 60^{\circ}$, with $|b| < 5^{\circ}$) but remains poor considering in average the whole Galactic disk  ($|l| < 80^{\circ}$, with $|b| < 5^{\circ}$).
}


{In order to have a deeper understanding of the discrepancy}, it is important to trace, for each line of sight (l.o.s.), which portion of the Galaxy the emission comes from. 
For this reason, in fig.~\ref{fig:Ring_Spectrum} we plot the relative contribution to the total $\pi^0$ emission for three reference l.o.s. {as a function of the Galactocentric 
distance, $R$}. At large values of the Galactic longitude $l$ (where the FB model gives a better fit) the emission is dominated by the local environment; instead, the closer to the center we look, the wider the relevant region gets, with the central rings contributing as much as $20\%$ for the Galactic center window (where the fit is worse and the data turn out to be significantly harder). 
{In the lower panel of fig.~\ref{fig:Ring_Spectrum}, we show the power-law spectral index of the $\pi^0$ component as a function of $R$; for the FB model, as expected, 
we find a constant value equal to the measured local proton spectral index.

Driven by these results,} {\ we argue that the FB model should be corrected in such a way to get a significantly harder propagated proton index for smaller values of $R$, and a value closer to the one inferred by Boron-to-Carbon ratio (B/C) and protons in the local region. We stress that, since in the sky windows where the emission is mostly local (at high longitude or high latitude), the contribution of IC and point sources to total emission is relevant, we never observe a $\gamma$-ray slope equal to the local $\pi^0$ slope.}

\section{Method}

We propose a propagation model based on the following three ingredients:

\begin{enumerate}[(i)] 

\item Bearing in mind the motivations outlined in the introduction, we drop the oversimplified assumption of constant diffusion, and we consider the possibility that the slope of the diffusion coefficient $\delta$ is a function of $R$. 

\item We allow for position-dependent convective effects;
the presence of a significant convective wind in the inner region of the Galaxy
is motivated by the X-ray observations by the ROSAT satellite \cite{Snowden:1997ze}, 
and may affect cosmic-ray propagation \cite{Gebauer:2009hk}.

\item We allow for  a larger value of ${\rm X}_{\rm CO}$ in the outer part of the Galaxy; this hypothesis stems from the existence of a gradient in metallicity across the Milky Way \cite{Cheng:2011xf}. The metallicity is a result of stellar and Galactic chemical evolution: it is higher towards the Galactic center, and decreases going outwards; since lower metallicities imply less dust shielding \citep{Israel1997}, it is reasonable to expect larger values of ${\rm X_{CO}}$ for increasing $R$. 

\end{enumerate}

For this purpose, we exploit the numerical packages {\tt DRAGON} \citep{Dragonweb,Evoli:2008dv} and {\tt GammaSky} (a dedicated code recently used in \citep{gradient:2012,Tavakoli2011,Cirelli2014} to simulate diffuse $\gamma$-ray maps). 

{

As a starting point, we consider the Kraichnan diffusion model defined in ref.~\cite{Evoli:2011id} (labeled KRA therein).\footnote{We checked that the same conclusions can be reached starting from 
the Kolmogorov and thick-halo diffusion models \citep{Evoli:2011id}.} As a first step, we modify $\delta$
introducing a functional dependence on $R$; as simplest and {\it a posteriori} sufficient guess, we consider $\delta(R) = A\,R + B$ with local normalization $\delta(R_{\odot}) = 0.5$, and -- to avoid unrealistically large values -- saturate it to $\delta(R > 11\,{\rm kpc}) = \delta(R = {11\,{\rm kpc}})$.
The free parameter $A$ is fixed by fitting the $\gamma$-ray data in the energy range $E_{\gamma}=[5-50]$ GeV;
to this purpose, we divide the Galactic disk $|b| < 5^{\circ}$, $|l| < 80^{\circ}$ in eight longitudinal windows of $10$ degrees each.

The energy spectra we obtain from this procedure 
correctly reproduce the measured slope in all the analyzed sky windows but overshoot the data at low energies, in particular for small values of $l$.
To tame this problem, in the inner region with $R < R_{\rm w}$, we allow for a strong convective wind with uniform gradient in the $z$-direction. 
We extract $R_{\rm w}$ and the intensity of the convective gradient 
by fitting the low-energy data with $E_{\gamma}< 1$ GeV.
Concerning the molecular hydrogen, we assume -- in units of $10^{20}$ cm$^{-2}$(K km s$^{-1}$)$^{-1}$ -- ${\rm X}_{\rm CO} = 1.9$ at $R < 7.5$ kpc, and ${\rm X}_{\rm CO} = 5$ at $R > 7.5$, in order to correctly match the normalization of the observed flux for $|l| > 50^\circ$. 

The last step of our method  consists in verifying {\it a posteriori} that the corrections described above do not spoil the local observables: we find that just a small tuning in the value of the normalization of the diffusion coefficient $D_0$ and in the source spectral index $\gamma$ are needed.

{ 
In particular, we checked protons (see Fig. \ref{fig:proton}), B/C (see Fig. \ref{fig:BC}), antiprotons (see Fig. \ref{fig:antiproton}), leptons, and $^{10}{\rm Be}/^9{\rm Be}$. 

\begin{figure}[!htb!]
\begin{center}
\centering
   \centering
   \includegraphics[scale=0.35]{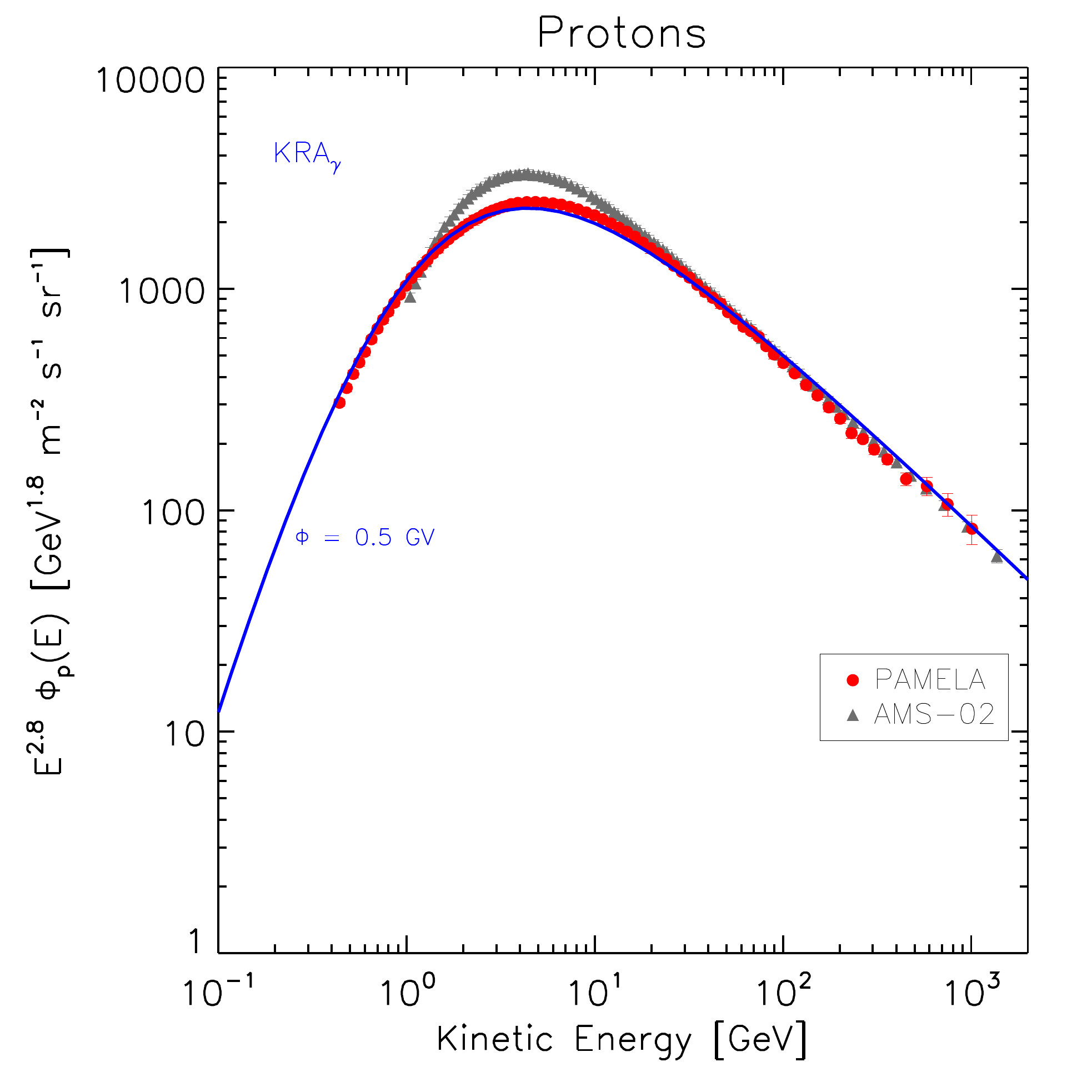}
\caption{\textit{
Comparison between the local proton flux in the KRA$_{\gamma}$ model and the corresponding experimental data. We use a fixed modulation potential of $500$ MV, and, 
in addition to the PAMELA data \cite{Adriani:2011cu}, 
we also show preliminary AMS-02 results \cite{AMSproton}.
}}
\label{fig:proton}
\end{center}
\end{figure}

\begin{figure}[!htb!]
\begin{center}
\centering
   \centering
   \includegraphics[scale=0.35]{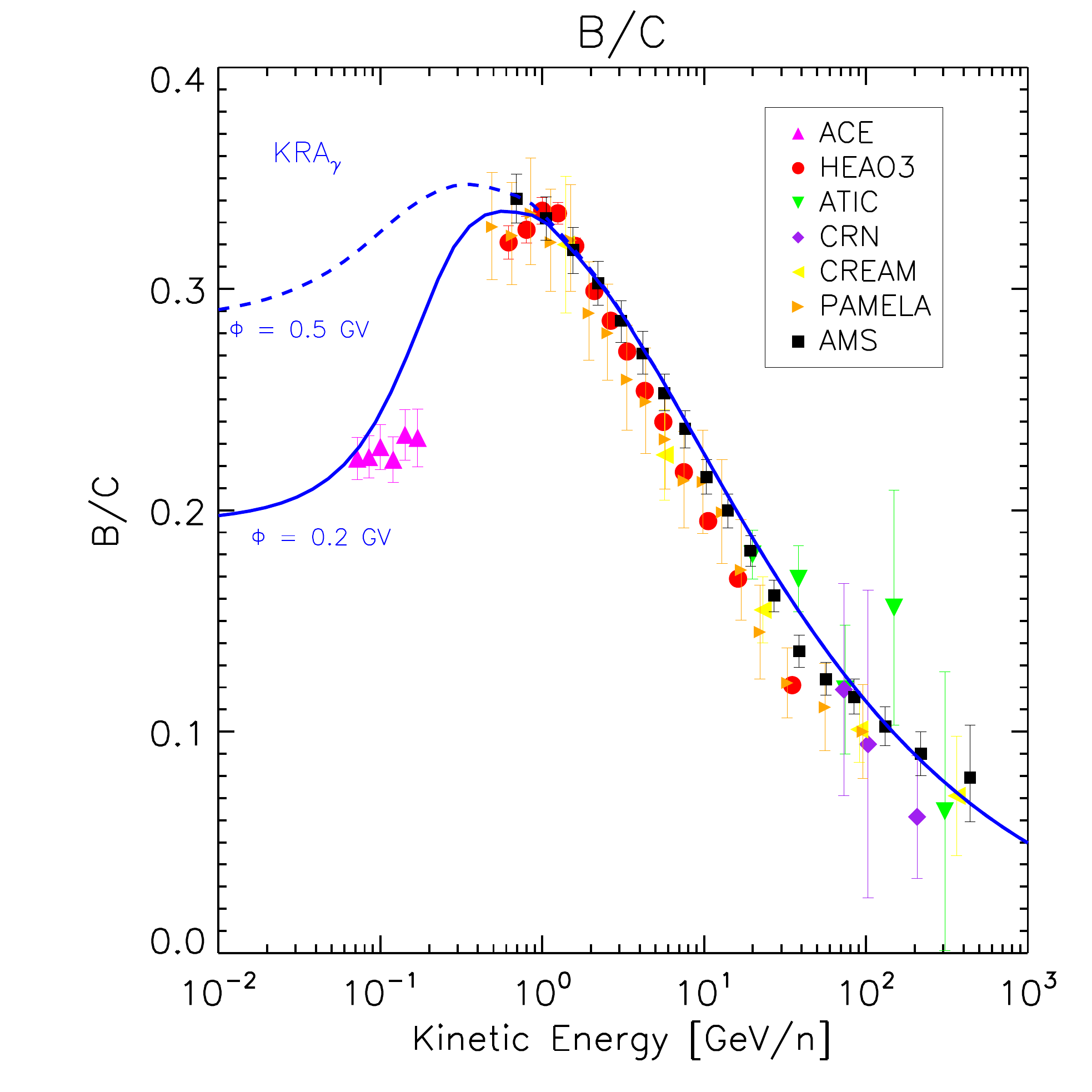}
\caption{\textit{
Comparison between the local B/C ratio in the KRA$_{\gamma}$ model and the corresponding experimental data. 
We show two different values for the modulation potential, $500$ MV (dashed line) and $200$ MV (solid line).
Data points refer to different experiments: ACE \cite{ACE}, HEAO-3 \cite{Engelmann:1990zz}, ATIC \cite{Panov:2007fe}, CRN \cite{CRN}, CREAM \cite{Ahn:2008my},
PAMELA \cite{Adriani:2014xoa} and AMS \cite{AMSBC}.
}}
\label{fig:BC}
\end{center}
\end{figure}

\begin{figure}[!htb!]
\begin{center}
\centering
   \centering
   \includegraphics[scale=0.35]{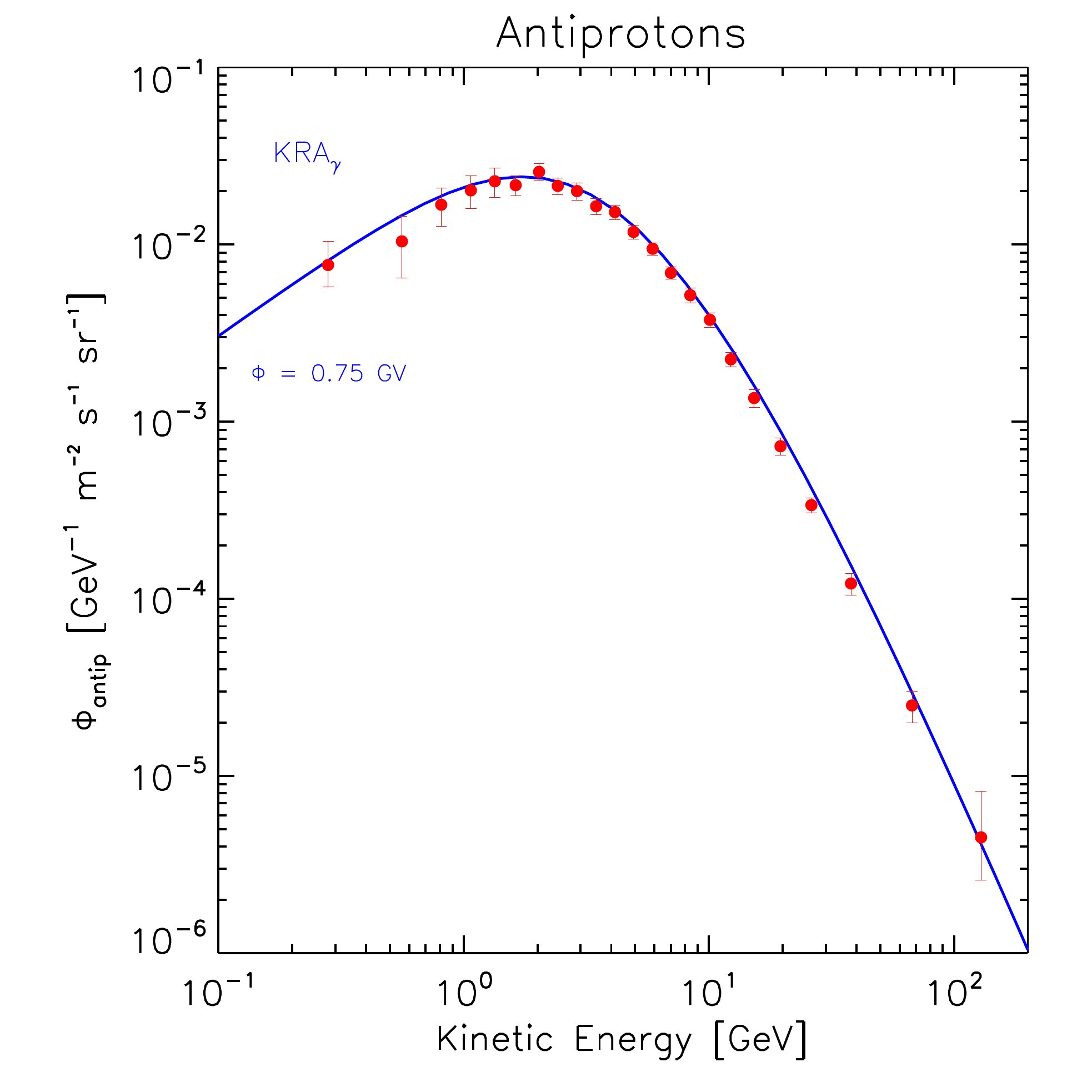}
\caption{\textit{
Comparison between the local antiproton flux in the KRA$_{\gamma}$ model and the corresponding PAMELA data \cite{Adriani:2012paa}. We use a fixed modulation potential of $750$ MV.
}}
\label{fig:antiproton}
\end{center}
\end{figure}

Concerning the Beryllium ratio, the compatibility between the observational evidence of strong winds in the inner Galaxy and the constraints from the radioactive isotopes may be a problem (see e.g. \cite{Gebauer:2009hk}). 
Nevertheless, in our case the Galactic wind is not present locally and therefore we have an acceptable agreement with the data.
}

All in all, we report the following best-fit values for the parameters described above: $A = 0.035$ kpc$^{-1}$, $R_{\rm w} = 6.5$ kpc, $dV/dz = 100$ km\,s$^{-1}$kpc$^{-1}$, $D_0 = 2.24\times 10^{28}$ cm$^2$s$^{-1}$, $\gamma = 2.35$. We label this model KRA$_{\gamma}$.
}

\section{Results.}

\begin{figure}[!htb!]
\centering
   \includegraphics[scale=0.45]{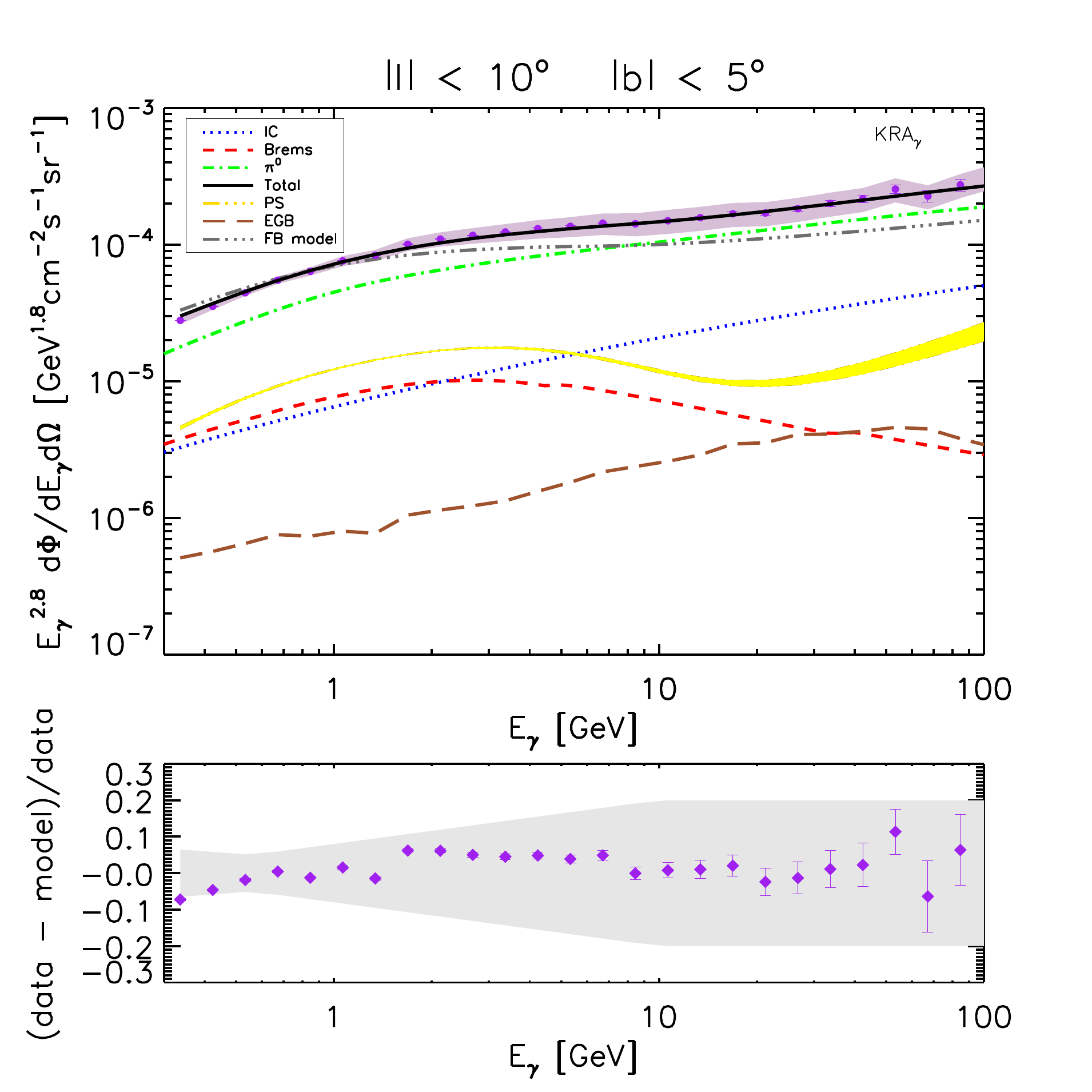}
\caption{\textit{
The same as in fig.~\ref{fig:Spectra_KRA_tuned_plane} but considering the window $|l| < 10^{\circ}$, $|b| < 5^{\circ}$.}}
\label{fig:KRA_Longitude_Window_0_10}
\end{figure}

\begin{figure}[!htb!]
\centering
   \includegraphics[scale=0.45]{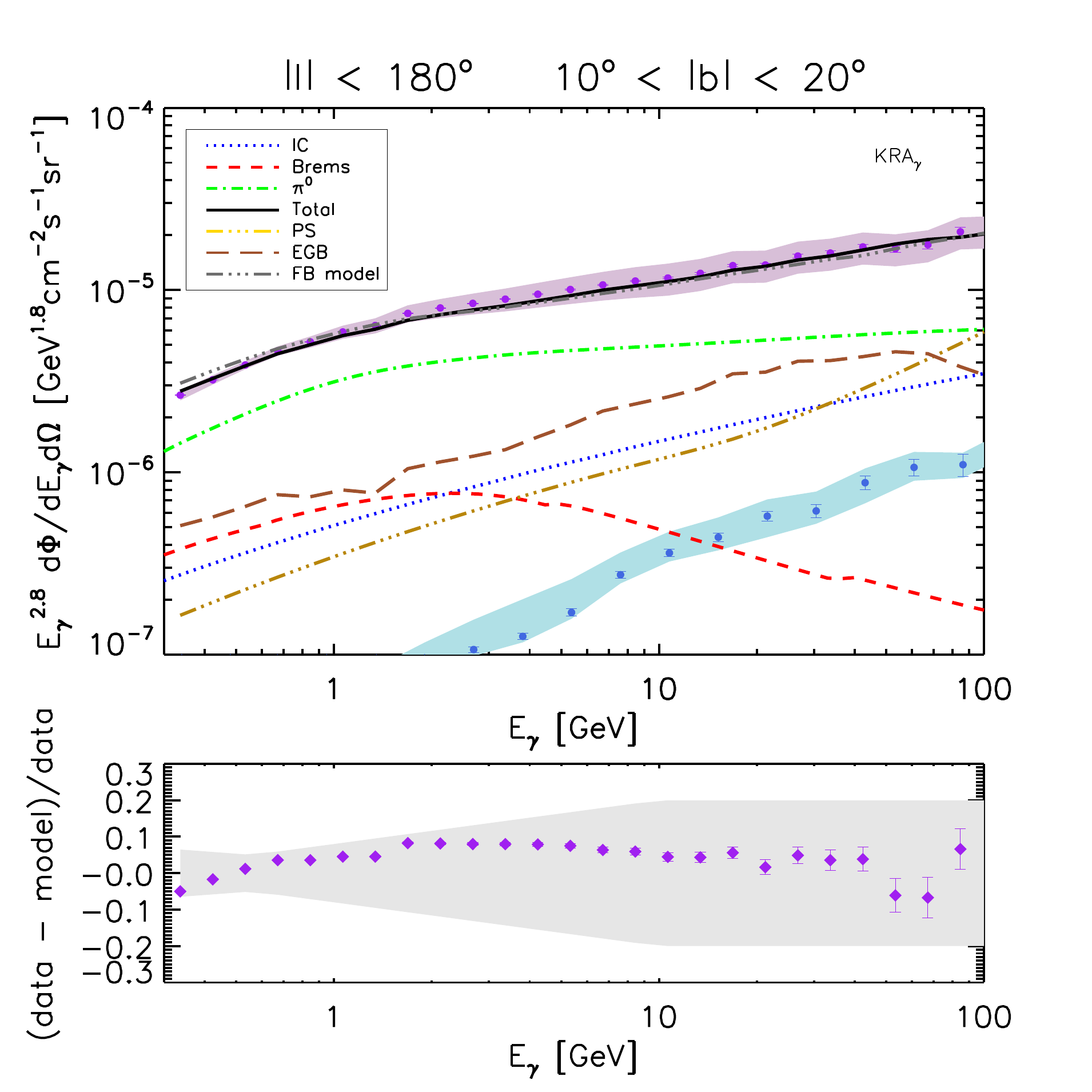}
\caption{\textit{The same as in fig.~\ref{fig:Spectra_KRA_tuned_plane} but considering the strip $|l| < 180^{\circ}$, $10^{\circ}< |b| <20^{\circ}$. 
The azure band represents the contribution of the Fermi bubbles according to ref.~\cite{Fermi-LAT:2014sfa}.}}
\label{fig:KRA_High_Latitude_10_20}
\end{figure}

We show in fig.~\ref{fig:Spectra_KRA_tuned_plane}, \ref{fig:KRA_Longitude_Window_0_10}, and \ref{fig:KRA_High_Latitude_10_20}, the $\gamma$-ray spectra for our ${\rm KRA_{\gamma}}$ model in three relevant sky windows: 
the Galactic disk, a small window focused on the Galactic center, and the mid-latitude strip with $|l| < 180^{\circ}$, $10^{\circ}< |b| <20^{\circ}$. 

\begin{figure}[!htb!]
\centering
   \includegraphics[scale = 0.45]{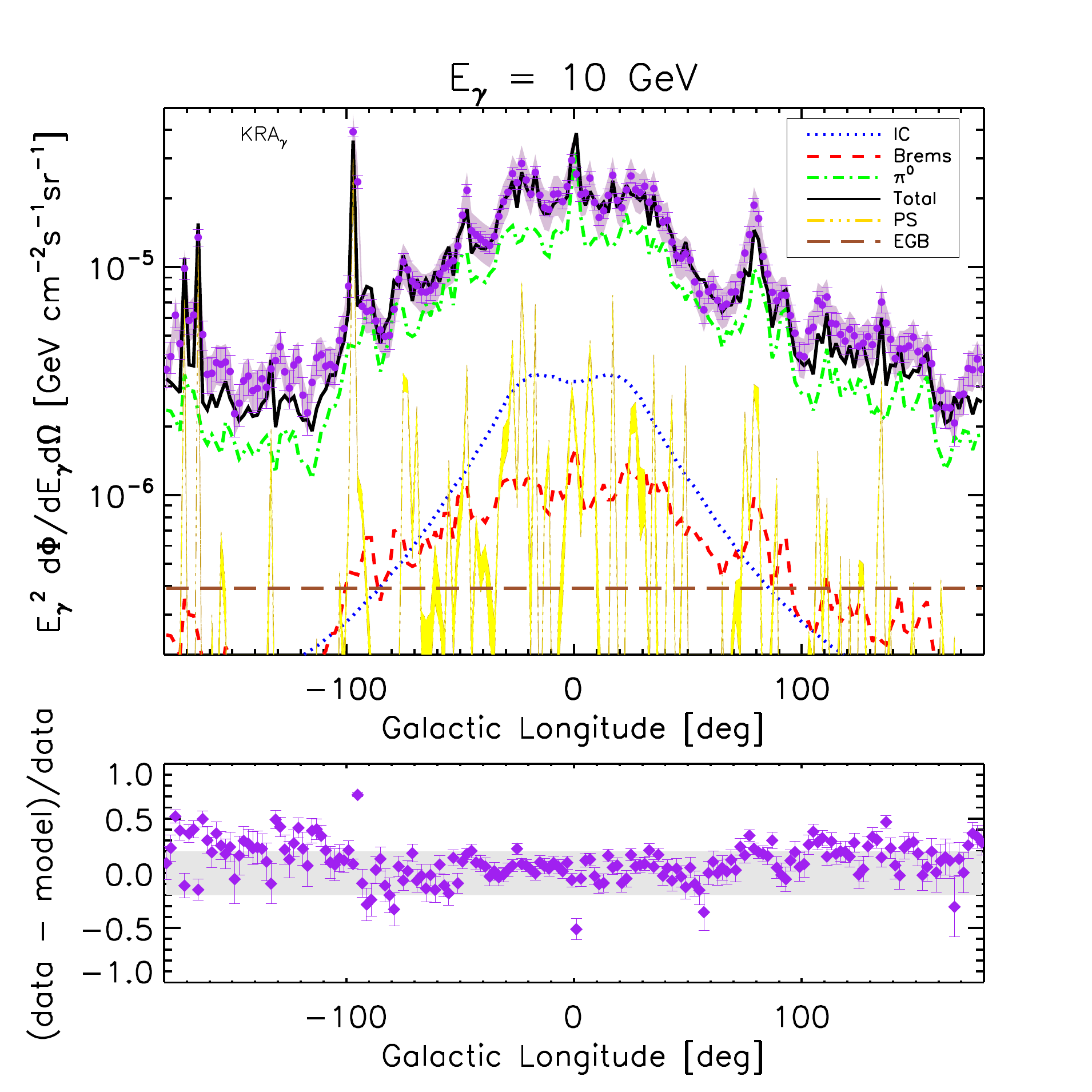}
\caption{\textit{Longitudinal profile at fixed energy $E_{\gamma} = 10$ GeV. We average in latitude over the interval $|b| < 5^{\circ}$.}}
\label{fig:Profile}
\end{figure}

In fig.~\ref{fig:Profile} we show the longitudinal profile. 
{ We remark that the model is not optimized for high longitudes ($|l| > 100^\circ$): this is the well-known {\it gradient problem}, and this discrepancy can be reabsorbed by a rescaling of the $\pi^0$ component --  motivated by the possible presence of neutral gas not traced by HI and CO emission lines -- a position-dependent normalization of the diffusion coefficient \citep{gradient:2012}, or an altered source term \cite{FermiQuadrant} with respect to the one we adopt \cite{Ferriere2001}.
A full-sky analysis based on a combined scenario with both a variable slope and normalization of the diffusion coefficient is far beyond the scope of this paper, and will be addressed in a future work.
}

In table~\ref{tab:chi2} we list the $\chi^2$ for our optimized model, showing a remarkable improvement with respect to the FB model. 


There are in principle alternative scenarios leading to tilted $\gamma$-ray fluxes, see e.g. \citep{FermiLAT:2012a,Berezhko2004,Erlykin2013,deBoer2014}. However: 

\begin{itemize}

\item Following ref.~\cite{Calore:2014oga}, { we find that a population of unresolved pulsars, consistent with the observed counterpart, gives an extra contribution to the total $\gamma$-ray flux more than one order of magnitude smaller than needed. }

{
\item Running a dedicated MC code where the analytical solution of the diffusion equation with the correct boundary is implemented, as described in \cite{Blasi:2011fi}, we simulate Supernova explosions with a reasonable rate $\simeq$ 3/century distributed according to the source term presented in \cite{Ferriere2001}. 

We fit each realization with a power-law.
We find that fluctuations in the proton spectrum due to the stocasticity of the sources never exceed -- even in the inner Galactic region -- the few percent level.
}

\item We test the possibility of an enhanced IC emission; we find that a rescaling of the ISRF by one order of magnitude, together with a factor of 10 decrease in the ${\rm X_{CO}}$, may solve the discrepancy. 

However, we discard this hypothesis since in this case the bulk of the $\gamma$-ray flux would have leptonic origin, in contrast with the obserbed correlation with the gas distribution as shown in 
fig.~\ref{fig:Profile}.

\end{itemize}

{
While the paper was undergoing the review process, the 4-year Point Source Catalog (3FGL) was released by the Fermi-LAT collaboration. We checked that our results are not affected by this update, given the subdominant role of point sources with respect to $\pi^0$ emission, especially in the windows near the Galactic center.
}

\section{Conclusions.}


We addressed the problem of modelling the $\gamma$-ray emissivity in the Galaxy from a new perspective. The aim was learning how the properties of CR diffusion change through the Galaxy. 
Our strategy consisted in developing a CR propagation model relaxing the assumption of homogeneous diffusion: we allowed $\delta$ to vary with the Galactocentric radius $R$. 
The main motivation is the discrepancy between the observed and predicted $\gamma$-ray slope: in particular, the models discussed in \cite{FermiLAT:2012a} underestimate the high-energy data in the Galactic plane region. 
Being the $\pi^0$ emission dominant at low latitudes, the $\gamma$-ray spectral index is determined by the proton spectrum; since the latter is well constrained by recent data, we assumed this tension to be a hint of a different diffusion regime taking place in the inner region of the Galaxy.  We adopted a minimal set of assumptions (linear variation of $\delta$, high convective regime for small $R$) and we found that our model reproduces the $\gamma$-ray data in many relevant windows of the sky within the systematic uncertainty. 
We achieved this result without relying on {\it ad hoc} tunings of astrophysical ingredients such as the gas distribution, the ${\rm X_{\rm CO}}$ conversion factor, the source distribution or the interstellar radiation field, and keeping a good agreement with locally measured CR spectra.
Remarkably, in the Galactic center window our residuals do not exceed the $10$\% level (see fig.~ \ref{fig:KRA_Longitude_Window_0_10}), which is comparable with the alleged Dark Matter signal reported in \citep{Morsellon,Goodenough2009,Hooperon:2014}. A more detailed analysis with focus on this region will be presented in a forthcoming work.

{\em Acknowledgments:} We thank I. Cholis, I. Gebauer, and D. Grasso for useful discussions and suggestions. 
The work of A.U. is supported by the ERC Advanced Grant n$^{\circ}$ $267985$, ``Electroweak Symmetry Breaking, Flavour and Dark Matter: One Solution for Three Mysteries" (DaMeSyFla).
P.U. and M.V. acknowledge partial support from the European Union FP7 ITN INVISIBLES (Marie Curie Actions, PITN-GA-2011-289442), and partial support by the research grant ``Theoretical Astroparticle
Physics'' number 2012CPPYP7 under the program PRIN 2012 funded by the Ministero dell'Istruzione, Universit\`a e della Ricerca (MIUR).
In this work the authors have used the publicly available {\tt Fermi-LAT} data and Fermi Tools
archived at \url{http://fermi.gsfc.nasa.gov/ssc/}.

\bibliographystyle{apsrev4-1}
\bibliography{bibliography}

\end{document}